\documentstyle[11pt,aaspp4]{article}
\def\Msun{~M_\odot}
\def\lsim{\raise0.3ex\hbox{$<$}\kern-0.75em{\lower0.65ex\hbox{$\sim$}}}
\def\gsim{\raise0.3ex\hbox{$>$}\kern-0.75em{\lower0.65ex\hbox{$\sim$}}}

\def\kms{\rm ~km~s^{-1}}

\def\ml{~\Msun ~\rm yr^{-1}}

\begin{document}

\title{RADIO SUPERNOVA SN 1998bw AND ITS RELATION TO GRB 980425}
\author{Zhi-Yun Li and Roger A. Chevalier}
\affil{Department of Astronomy, University of Virginia, P.O. Box 3818}
\affil{Charlottesville, VA 22903; zl4h@virginia.edu, rac5x@virginia.edu}

\begin{center}
version: \today
\end{center}

\begin{abstract}

SN 1998bw is an unusual Type Ic supernova that may be associated with 
the $\gamma$-ray burst GRB 980425. We use a
synchrotron self-absorption model for its radio emission
to deduce that the synchrotron-emitting gas is 
expanding into a circumstellar medium of approximately $r^{-2}$ density 
profile, at a speed comparable to the speed of light. 
We assume that the efficiencies of production of relativistic electrons
and magnetic field are constant through the evolution.
The circumstellar density is consistent with that expected around
the massive star core thought to be the progenitor of SN 1998bw.
The  explosion
energy in material moving with velocity $>0.5c$ 
is   $\sim 10^{49}- 3\times 10^{50}$ ergs, with some 
preference for the high values. 
The rise in the radio light curves observed at days $20-40$ is inferred
to be the result of a rise in the energy of the blast wave 
 by a factor $\sim 2.5$. 
Interaction with a jump in the ambient density is not
consistent with the observed evolution.
We infer  that the boost in energy is from a shell of matter from
the explosion that catches up with the decelerating shock front.
Both the high explosion energy and the nature of the energy input to
the blast wave are difficult to reconcile with
energy input from the shock-accelerated high
velocity ejecta from a supernova.
The implication is that there is irregular energy input from a central
engine, which is the type of model invoked for normal $\gamma$-ray bursts.
The link between SN 1998bw and GRB 980425 is thus strengthened.

\end{abstract}


\section{INTRODUCTION}

SN 1998bw was discovered in an optical search of the error circle of
the gamma-ray burst (GRB) source GRB 980425 (Galama et al. 1998b).
The light curve of the supernova indicates an explosion time that
is consistent with the time of GRB 980425 and Galama et al. (1998b)
 place the probability of a chance coincidence of the
supernova and GRB at 10$^{-4}$.
The supernova spectrum implied a Type Ic event, but one with an unusually
high luminosity and high material velocities.
Modeling of the light curve of SN 1998bw suggested an explosion energy
of $(2-3)\times 10^{52}$ ergs (Iwamoto et al. 1998; Woosley, Eastman,
\& Schmidt 1999), more than an order of magnitude higher than a
typical supernova energy.
Kulkarni et al. (1998) discovered a variable radio source coincident
with the optical supernova.
Although a number of Type Ib/c supernovae have been found to be
radio sources (Weiler et al. 1999), SN 1998bw was more 
luminous by $\sim 10^2$ at its peak.
The unusual, energetic nature of SN 1998bw would seem to further
strengthen its  association with GRB 980425.
Observations with
{\it BeppoSAX} showed two variable X-ray sources in
the error box of GRB 980425, one of which
is coincident with SN 1998bw (Pian et al. 1999;
Galama et al. 1998b).
The other source might have been the X-ray afterglow of the GRB 980425,
but the evidence for this is not compelling.

The aim of this paper is to develop a model for RSN (radio supernova) 1998bw 
 and to relate its 
properties to
 those of SN 1998bw and GRB 980425.
Here, we use the term RSN 1998bw 
primarily as an abbreviation for ``the radio
emission from SN 1998bw.''
Previously, Kulkarni et al. (1998) argued that the shock velocity 
in SN 1998bw is
relativistic based on the high brightness
temperature, on a synchrotron self-absorption interpretation of the early
evolution, and on scintillation results.
They suggested that an early phase of this shock wave produced GRB 980425.
Waxman \& Loeb (1999)  modeled the spectrum of RSN 1998bw and proposed that
the emission is from a thermal distribution of electrons with
a shock velocity of $0.3c$.
They use this result to infer that the supernova may be unrelated
to GRB 980425.
Our model uses a more complete data set and includes a 
relativistic dynamical model.
Our conclusions are closer to those of Kulkarni et al. (1998)
than to those of Waxman \& Loeb (1999).

The plan of the paper is as follows.
The properties of RSN 1998bw are compared to those of other radio supernovae
in \S~2.
In \S~3, we present a nonrelativistic model for RSN 1998bw.
We assume that a power law electron spectrum is produced in the shock front.
We treat dynamical and relativistic effects in \S~4 and thus
restrict the range of possible models.
Approximate scaling laws for the range of models are discussed in \S~5.
Our model for RSN 1998bw is placed in the context of the properties of
SN 1998bw and GRB 980425 in \S~6.
We present a final discussion and conclusions in \S~7.

\section{COMPARISON TO RADIO SUPERNOVAE}

There is a growing data set on radio supernovae of Types Ib/c and II
(Weiler et al. 1998, 1999 and references therein).
The overall radio evolution can be described as an early phase
when optical depth effects are important followed by a transition
to a phase with an optically thin power law spectrum 
(flux $\propto \nu^{-(p-1)/2}$).
The power law index for the electron
spectrum, $p$, has not been observed to vary in individual
supernovae, although it does vary in the range $2.1-3.2$ for the
group of radio supernovae (Van Dyk et al. 1994a).
The time evolution in the optically thin regime can be described
by a power law decrease (flux $\propto t^{-w}$), with $w$ ranging
from 0.66 (SN 1980K; Weiler et al. 1992) to 1.6 (SN 1983N).

The evolution of the fluxes from RSN 1998bw (Kulkarni et al. 1998; 
Wieringa, Kulkarni, \& Frail
1999; Fig. 1)  and the evolution of the spectral indices
(Fig. 2) show that the description of radio supernovae in the
optically thin regime  roughly applies to RSN 1998bw
for ages $t\gsim 40$ days.
The decline is approximately a power law in time with $w=1.6$ (see
also Iwamoto 1999).
The electron spectral index shows some variation but generally is in the
range $p=2.2-2.6$.
There are deviations from the late power law behavior, but the deviations
are comparable to those found in other radio supernovae.
SN 1998bw belongs to the class of Type Ic supernovae
(Galama et al. 1998b).
Because Type Ib and Type Ic supernovae have similar properties and differ
from each other only in the presence or absence of optical He lines
near maximum light, we compare SN 1998bw to both of these types.
The parameters for the late evolution of radio supernovae of these
types are SN 1983N (Ib): $p=3.0$, $w=1.6$;
SN 1984L (Ib): $p=3.0$, $w=1.5$;
SN 1990B (Ic): $p=3.2$, $w=1.3$ (Van Dyk et al. 1994a).
Radio emission has also been observed from the Type Ic SN 1994I,
but the results have not been fully published.
Observations at 15 and 88 GHz indicate an optically thin spectral
index of $p=2.8\pm 0.2$ (Wink 1994).
The radio spectrum of RSN 1998bw is flatter than those of other Type Ib/c
supernovae and its decline rate is at the high end of those observed.

The previously observed Type Ib/c supernovae have similar radio 
luminosities, within a factor of $2-3$.
Fig. 3 (based on Fig. 4 of Chevalier 1998) shows the times and peak luminosities at 6 cm
of radio supernovae; the Type Ib/c supernovae occupy a small area on
this plot.
SN 1998bw is extraordinary in that its radio luminosity is $\sim 10^2$ higher than
that of the other Type Ib/c supernovae.
It remains $\sim 10^2$ stronger than SN 1983N over the full observed
period of 250 days.
Kulkarni et al. (1998) used the high brightness temperature of the
source to argue for relativistic expansion.
The straight lines in Fig. 3 are from the theory described in Chevalier (1998)
which assumes that optically thick phase of the supernova is due to
synchrotron self-absorption.
The electron spectrum is assumed to be a power law ($p=2.5$ in Fig. 3)
which extends down to transrelativistic energies.
An expression for the radius of the emitting region can then be obtained
which depends on the factor (magnetic energy density/relativistic electron
energy density)$^{1/(2p+13)}$.
In Fig. 3, the factor is taken to be unity; the weak dependence on the
energy density ratio provides  justification for the assumption.
The total energy in fields and particles goes up if the system is far from
equilibrium.
The position of SN 1998bw in Fig. 3 is suggestive of 
relativistic motion, which means that the nonrelativistic theory used for
Fig. 3 breaks down.

In addition to its high luminosity, RSN 1998bw is also unusual in its
 evolution leading to the late power law decline.
Fig. 1 shows that the power law decline began at high  frequencies
(4.80 and 8.64 GHz) on day 12 and continued through day 22.
Between days 22 and 32, there was a rapid increase at these 
frequencies which
raised the level of the power law decay by a factor of $\sim 2.2$.
This behavior has not been previously observed in the light curves
of radio supernovae,
although SN 1987A has shown a strong rise in its radio flux 
(Ball et al. 1995) which
can be attributed to interaction with the dense wind from a previous
evolutionary phase.
The observed light curves generally do show fluctuations about smooth curves
(Weiler et al. 1999) similar to that seen in the late power law evolution
of RSN 1998bw.
%but they have not shown a systematic shift in level.
The radio data on many supernovae are fragmentary, but there are excellent
data at 5 frequencies on SN 1993J (Van Dyk et al. 1994b; Weiler et al. 1999).
The radio evolution is smooth and can be modeled by shock-accelerated
supernova ejecta running into a $\rho\propto r^{-2}$ circumstellar medium
(Fransson \& Bj\"ornsson 1998).
Another anomaly in the early evolution of RSN 1998bw is 
the rise at 4.80 and 8.64 GHz before
day 10.
The rise occurs with little change in spectral index, which implies that
it is not due to a reduction in optical depth, as is the case in
other radio supernovae.

\section{A MODEL FOR RSN 1998bw}

The properties of RSN 1998bw sufficiently resemble those of other radio 
supernovae to initially consider a model similar to those applied to
radio supernovae.
The standard model for these objects involves synchrotron
emission from relativistic electrons in a spherical shocked region
between the supernova and a surrounding circumstellar wind (Chevalier 1982).
The early optically thick phase can be dominated by either free-free
absorption or synchrotron self-absorption, but the high velocities
and low circumstellar densities found in Type Ib/c supernovae implies
that synchrotron self-absorption is the dominant mechanism in 
these objects (Slysh 1990; Chevalier 1998).
The same arguments apply to the Type Ic RSN 1998bw 
(see also Kulkarni et al. 1998).
A significant new feature for RSN 1998bw is that the high shock
velocity inferred for this radio supernova implies that all the 
postshock electrons are accelerated to relativistic energies.
As in studies of GRB afterglows (Waxman 1997; Wijers, Rees, \& M\'esz\'aros
1997), we assume that the
 distribution of relativistic electrons with 
 the Lorentz factor $\gamma$ takes a power law form with
the number density given by
\begin{equation}
 n_e(\gamma)\ d\gamma=C\ \gamma^{-p}\ d\gamma, 
\end{equation} 
above a lower limit $\gamma_m$, which is determined by the shock velocity.
Support for the assumption in the GRB case can be found in the afterglow
emission from GRB 970508 (Galama et al. 1998a).
This distribution function distinguishes our model for RSN 1998bw from that
of Waxman \& Loeb (1999), who assumed a thermal (essentially monoenergetic)
electron distribution.

Relativistic electrons moving with a Lorentz factor of $\gamma$ emit
synchrotron radiation of a characteristic frequency 
\begin{equation}
\nu_c(\gamma)={3\gamma^2 q B_\perp\over 4\pi mc},
\end{equation}
where $q$ and $m$ are the charge and the mass of the electron, and 
$B_\perp$ the strength of the component of magnetic field perpendicular 
to the electron velocity. 
 In the simplest case where the frequency
of interest $\nu>>\nu_m=\nu_c(\gamma_m)$, the 
synchrotron spectrum has a $\nu^{5/2}$ spectral dependence in the 
optically thick region and a $\nu^{-(p-1)/2}$ dependence in the 
optically thin region (e.g., Rybicki \& Lightman 1979). 
This simple model spectrum fails 
to reproduce two properties of the observed radio spectrum: (1) at the 
earliest times when the radio emission is most probably optically thick 
at the two highest frequencies (3 cm and 6 cm), the spectral index 
between them is closer to $2$ than to $5/2$, as noted by Kulkarni et al.
(1998) and Waxman \& Loeb (1999); and (2) the observed spectrum is 
significantly broader than the above simple spectrum. These two
discrepancies lead us to believe that the characteristic frequency 
$\nu_m$ is important in shaping the observed spectrum. We explore 
this possibility and try to infer the physical properties of the 
radio emitting plasma from $\nu_m$ and other parameters that 
determine the spectrum.
A self-absorbed spectral index of 2 (rather than 5/2) for radiation
produced by electrons accelerated by a relativistic shock was predicted by
Katz (1994).

In this section, we restrict our discussion to emission 
from a nonrelativistically moving
medium. It is well known that, in a synchrotron self-absorption model, 
the flux at a given frequency $\nu$ is proportional to
\begin{equation}
f_\nu\propto\nu^{5/2} \left(1-e^{-\tau}\right) 
	{\int^{x_m}_0 F(x) x^{(p-3)/2} dx\over \int^{x_m}_0
	F(x) x^{(p-2)/2} dx} ,
\end{equation}
where $x_m=\nu/\nu_m$, $F(x)$ is a known function of $x$ (shown, e.g., on 
page 179 of Rybicki \& Lightman 1979), and $\tau$ is the optical depth
of synchrotron self-absorption. From the standard theory, we have 
\begin{equation}
\tau\propto \nu^{-(2+p/2)} \int^{x_m}_0 F(x) x^{(p-2)/2} dx.
\end{equation}
To simplify expressions, we define two auxiliary functions
\begin{equation}
F_1(\nu,\nu_m,p)=\int^{x_m}_0 F(x) x^{(p-3)/2} dx; \ \ \ 
F_2(\nu,\nu_m,p)=\int^{x_m}_0 F(x) x^{(p-2)/2} dx,
\end{equation}
and denote the flux and the optical depth at a reference frequency
$\nu_0=1$ GHz by $f_0$ and $\tau_0$ respectively. The flux at 
other frequencies can then be obtained from
\begin{equation}
{f_\nu\over f_0}={F_1(\nu)\over F_1(\nu_0)}{F_2(\nu_0)\over F_2(\nu)}
	\left({\nu\over\nu_0}\right)^{5/2}\left\{1- \exp\left[-\tau_0
	\left({\nu_0
	\over\nu}\right)^{2+p/2}{F_2(\nu)\over F_2(\nu_0)}\right]
	\right\},
\end{equation}
where the dependence of $F_1$ and $F_2$ on $\nu_m$ and $p$ is suppressed
for clarity. Note that the relative flux $f_\nu/f_0$ depends on three 
parameters: $\tau_0$ (the optical depth at $\nu_0=1$ GHz), $\nu_m$ 
(the characteristic frequency corresponding to the lowest electron
energy) and $p$ (the electron energy distribution index). By fitting 
the shape of radio spectrum at each day, we hope to determine these 
three parameters as a function of time.

The shape of the observed spectrum is specified by three spectral indices, 
between 20 cm/13 cm, 13 cm/6 cm, and 6 cm/3 cm (Fig. 2). 
 In principle, one can
invert equation (6) to determine the parameters $\tau_0$, $\nu_m$ and 
$p$ uniquely from the three spectral indices. In practice, we adopt
simple power-law distributions for these parameters as a function of 
time and compute and match the three spectral indices to the observed
values as closely as possible. Through trial and error, we find that 
reasonable fits are obtained 
for
\begin{equation}
\tau_0=25\left({t\over 10\ \ \hbox{\rm days}}\right)^{-2}; \ \ \ 
\nu_m=5.6\left({t\over 10\ \ \hbox{\rm days}}\right)^{-1}{\rm GHz}; \ \ \ 
p=2.5.
\end{equation}
The fits are shown in Figure~2. 
We should point out that the spectral indices
between 6 cm and 3 cm on day 3 and day 4 are not reproduced by our
fits, indicating that processes other than synchrotron self-absorption 
might be important at such early times or that the simple power law
evolution breaks down. 
In addition, both $\nu_m$ and the self-absorption turnover frequency
have moved below the observed frequency range by day 50 so these
parameters are not constrained by the observations at later times.
Over the time period that they are important
for the observed radio emission, $\nu_m$ and the self-absorption
 turnover frequency are close to each other.

To convert the fitted optical depth $\tau_0$, characteristic frequency 
$\nu_m$, and index $p$ into physical quantities of the synchrotron 
emitting gas, we need to have a specific model for its geometry. Here
we adopt the simplest model of a thin, uniform, spherical shell of 
radius $r$ and thickness $\xi$ ($< 1$) times the radius. For a
given value of $p$, the optical depth, the frequency $\nu_m$ and 
the observed flux at a given frequency (say, 6 cm), we
determine three physical quantities, which we choose  to be 
the total energy $E$, the radius $r$ and the electron density $n_e$.

For a power-law distribution of electrons given by equation (2), we 
have from the standard theory  
\begin{equation}
\tau= {p+2\over 2\pi m\nu^2}{\sqrt{3}q^3\over 2 m c^2}
	\left({4\pi m c\nu\over 3 q}\right)^{-p/2}F_2(\nu)
	C B_\perp^{(p+2)/2}(\xi r),
\end{equation}
taking into account the effect of spherical geometry
approximately (Chevalier 1998), and 
\begin{equation}
\nu_m={3\gamma_m^2 q B_\perp\over 4\pi mc},
\end{equation}
\begin{equation}
f_\nu={\pi r^2\over d^2} {2m\nu^2\over p+2}\left({4\pi mc\nu\over
	3 q B_\perp}\right)^{1/2}{F_1(\nu)\over F_2(\nu)}
	\left(1-e^{-\tau}\right),
\end{equation}
where the electron distribution coefficient $C$ is related to the electron
energy density $U_e$ by
\begin{equation}
C=(p-2){U_e\over mc^2}\gamma_m^{p-2},
\end{equation}
and the minimum Lorentz factor $\gamma_m$ is given, from equation (9), 
by
\begin{equation}
\gamma_m=\left({4\pi m c \nu_m\over 3 q B_\perp}\right)^{1/2}.
\end{equation}
Making the usual assumption that the energy densities of electrons and 
magnetic fields are proportional to the total energy density $U$, we 
have $U_e=\epsilon_e U$ and $B_\perp=(8\pi\epsilon_b U)^{1/2}$
\footnote{ Note that the magnetic energy fraction defined in this 
paper, $\epsilon_b=B_\perp^2/(8\pi U)$, is one-third the more
conventionally used fraction, $\epsilon_B=B_{\rm tot}^2/(8\pi U)$ 
(where $B_{\rm tot}$ is the total field strength), for an isotropic
electron distribution.}.
It turns out only two physical quantities enter equations (8) and (10)
for the optical depth and the flux. They are the radius $r$ of the
emitting region and the total energy density $U$, or alternatively 
the total energy $E=4\pi \xi r^3 U$. Straightforward but tedious 
algebraic manipulations show that 
$$
 E= 5.1\times 10^{48} 
	\left({0.1\over\epsilon_b}\right)^{6/17}
	\left({0.1\over\epsilon_e}\right)^{11/17}
	\left({\xi\over 0.1}\right)^{6/17}
	\left({d\over\ 40\ \hbox{\rm Mpc}}\right)^{40/17}
$$
$$
	\left({f_\nu\over 20\ \hbox{\rm mJy}}\right)^{20/17}
	\left({p+2\over 5}\right)^{9/17}\left({1\over p-2}\right)^{11/17}
	\left({\nu\over 1\ \hbox{\rm GHz}}\right)^{(11p-56)/34}
$$
\begin{equation}
	\left({\nu_m\over 1\ \hbox{\rm GHz}}\right)^{11(2-p)/34}
	{F_2(\nu)^{9/17}\tau^{11/17}\over
	F_1(\nu)^{20/17}\left(1-e^{-\tau}\right)^{20/17}}\ \ \hbox{\rm erg}
\end{equation}
and 
$$
 r=1.9\times 10^{17} 
	\left({\epsilon_b\over 0.1}\right)^{1/17}
	\left({0.1\over\epsilon_e}\right)^{1/17}
	\left({0.1\over \xi}\right)^{1/17}
	\left({d\over 40\ \hbox{\rm Mpc}}\right)^{16/17}
$$
$$
	\left({f_\nu\over 20\ \hbox{\rm mJy}}\right)^{8/17}
	\left({p+2\over 5}\right)^{7/17}\left({1\over p-2}\right)^{1/17}
	\left({\nu\over 1\ \hbox{\rm GHz}}\right)^{(p-36)/34}
$$
\begin{equation}
	\left({\nu_m\over 1\ \hbox{\rm GHz}}\right)^{(2-p)/34}
	{F_2(\nu)^{7/17}\tau^{1/17}\over
	F_1(\nu)^{8/17}\left(1-e^{-\tau}\right)^{8/17}}\ \ \hbox{\rm cm}.
\end{equation}
In these expressions, the quantities $p$, $\nu_m$ and $\tau$ (as well as 
$F_1(\nu)$ and $F_2(\nu)$ implicitly) are given by equations (7) and 
(4), obtained from fitting the spectral shape of the radio emission. 
The flux $f_\nu$
is given by the observed radio light curves. We choose the smoothest
curve of $6$ cm, and fit it with an analytical expression, as shown
in Figure~1. Other model curves on the same figure are obtained using 
the fitted parameters in equation (7). They match the data reasonably
well.
The weak dependence of $r$ on $\epsilon_b/\epsilon_e$ is comparable
to that found for normal radio supernovae (Chevalier 1998).

    From the fitted flux at $6$ cm and equations (13)-(14), we obtain 
the time evolution of the total internal energy and the radius of
synchrotron emitting gas. For a typical set of parameters $\epsilon_b
=\epsilon_e=\xi=0.1$ and $d=40$ Mpc, we find a total energy between
about $0.4\times 10^{49}$ -- $1.2\times 10^{49}$ erg  and a 
radius between about 
$0.5\times 10^{17}$ -- $2.5\times 10^{17}$ cm in the time interval 
from day 12 to day 80, as shown in panels 
(a)-(b) of Figure~4. For comparison, a straight line with a slope 
equal to the speed of light is drawn in panel (b), where the radius
is plotted as a function of time. 
The average 
speed in this time interval is close to the speed of light, although 
the instantaneous value could be somewhat different. 

We can also infer the number density of the emitting electrons, $n_e$. 
  From equation (2), we have 
\begin{equation}
n_e=\int_{\gamma_m}^\infty n_e(\gamma)\ d\gamma={p-2\over p-1}{\epsilon_e
	\over m c^2}{U\over \gamma_m}.
\end{equation}
A straightforward substitution yields 
$$
n_e=1.0 
	\left({0.1\over\epsilon_b}\right)^{7/17}
	\left({\epsilon_e\over 0.1}\right)^{7/17}
	\left({0.1\over \xi}\right)^{10/17}
	\left({40\ \hbox{\rm Mpc}\over d}\right)^{10/17}
$$
$$
	\left({20\ \hbox{\rm mJy}\over F_\nu}\right)^{5/17}
	\left({5\over p+2}\right)^{15/17}{(p-2)^{7/17}
	\over p-1}\left({\nu\over 1\ \hbox{\rm GHz}}\right)^{(10p+65)/34}
$$
\begin{equation}
	\left({\nu_m\over 1\ \hbox{\rm GHz}}\right)^{(3-10p)/34}
	{F_1(\nu)^{5/17}\tau^{10/17}\left(1-e^{-\tau}\right)
	^{5/17}\over F_2(\nu)^{15/17}}\ \ \hbox{\rm cm}^{-3}.
\end{equation}
This quantity is shown as a function of radius in panel (c) of Figure~4. 
Clearly, the number density of the emitting electrons drops off with 
radius approximately as $r^{-2}$. Since the density in the emitting 
region is related to the density in the ambient medium
(which supplies the emitting gas with matter through a strong shock)
by a simple compression factor, we conclude that the ambient medium
must have a $r^{-2}$ density profile as well. Such a profile arises 
naturally in a stellar wind of constant mass loss rate and constant 
velocity from the progenitor star of the supernova. We can estimate the 
mass loss rate in the wind through
\begin{equation}
{\dot M}_w=4\pi\rho_w V_w r^2= 8\pi \xi n_e m_p r^2 V_w,
\end{equation}
where the subscript ``$w$'' stands for the wind.  We have adopted a 
nucleon-to-electron number density ratio of two, appropriate for the 
predominantly helium (and perhaps carbon/oxygen) wind of a Wolf-Rayet 
star, which is the likely progenitor of SN 1998bw (see \S~6 for
discussion).  
We plot the inferred mass flux as a function of radius in panel (d)
of Figure~4. For the typical parameters chosen, we have a stellar
 mass loss rate of order $2.5\times 10^{-7} (V_w/ 10^3\ \hbox{\rm 
km\ s}^{-1})\ml$.  
The dependence on the  parameters $\epsilon_b$ and $\epsilon_e$
can be found from equation (16).

\section{DYNAMICAL AND RELATIVISTIC EFFECTS}

So far, we have worked directly with the radio data and inferred from it 
properties of the emitting region using a simple radiation model. Now 
we use these results as a guide and construct a more detailed model 
taking into account the dynamics of the emitting matter and 
relativity - two principal effects neglected in the above approach. 

We envision an instantaneous release of a large amount of energy $E_0$ 
in a $r^{-2}$-density external medium. The energy released  in the 
medium drives a blast wave whose dynamics can be deduced approximately from 
the following considerations. In the frame at rest with respect
to the origin of the explosion (and the observer), let the shocked 
external material 
be distributed uniformly in a shell of thickness $\Delta R_s$ (much 
less than the outer radius of the shell, the shock radius $R_s$). The
physical quantities inside the shell are related to those of the 
external medium by a set of shock-jump conditions, which
 are given by Blandford \&
McKee (1976) for an arbitrary strong shock.
 Let $\gamma$ be the bulk 
Lorentz factor of the shocked medium in the shell and $\Gamma$ that 
of the shock front. The  jump conditions are 
$$
U^\prime={\eta\gamma+1\over \eta-1}(\gamma-1)nm_pc^2,
\eqno(18)
$$
$$
n^\prime={\eta\gamma+1\over\eta-1} n,
\eqno(19)
$$
where $U^\prime$ and $n^\prime$ are the energy and nucleon 
number densities
of the shell in its comoving frame, and $\eta$ is the adiabatic 
index, which equals $4/3$ for ultrarelativistic shocks and $5/3$ 
for nonrelativistic shocks. A simple interpolation between these 
two limits, 
$$
\eta={4\gamma+1\over 3\gamma},
\eqno(20)
$$
should be valid approximately for transrelativistic shocks as well 
(Huang, Dai \& Lu 1998). The symbol $m_p$ denotes the mass of 
protons while $n$ is the  nucleon number density of the 
external medium, 
$$
n={n_0 r_0^2\over r^2},
\eqno(21)
$$
where $n_0$ and $r_0$ are constants. The Lorentz factor of the shock 
front is given by
$$
\Gamma^2={(\gamma+1)[\eta(\gamma-1)+1]^2\over \eta(2-\eta)(\gamma-1)+2},
\eqno(22)
$$
which depends on the shell Lorentz factor $\gamma$.

The shell Lorentz factor $\gamma$ is determined by the conservation of
mass
$$
4\pi R_s^2\Delta R_s {\hat n} = 4\pi n_0 r_0^2 R_s
\eqno(23)
$$
and the conservation of energy
$$
4\pi R_s^2\Delta R_s {\hat U} = E_0,
\eqno(24)
$$
where the energy density ${\hat U}$ and the number density ${\hat n}$
of the shell in the frame at rest with respect to the origin are 
related to those in the comoving frame by the transformation
$$
{\hat U}={4\gamma^2-1\over 3} U^\prime, \ \ \ \ {\hat n}=\gamma n^\prime.
\eqno(25)
$$
The transformation of the energy density is not exact in the
transrelativistic regime, but is intended to have the correct ultrarelativistic
and non-relativistic limits.
Combining equations (23)-(25) we obtain an algebraic equation for $\gamma$
$$
{(\gamma-1)(4\gamma^2-1)\over 3\gamma}={E_0\over 4\pi m_pc^2 n_0 r_0^2 R_s}
\eqno(26)
$$
in terms of the shock radius $R_s$. The shock radius itself evolves 
according to 
$$
{dR_s\over dt}=c\sqrt{1-\Gamma^{-2}},
\eqno(27)
$$
where $t$ is the time since the explosion in the rest frame of the origin. 
This completes our discussion of the blast wave dynamics.

To properly calculate emission from transrelativistic blast waves, one 
needs to take into account the finite light-travel time (Rees 1967).  
To facilitate discussion, let us adopt a cylindrical coordinate 
system ($R, \phi, z$), with the $z$-axis pointing from the origin of 
the blast wave toward the observer (see Figure~5). Let $T$ be the 
time in the observer's frame. A light pulse emitted at a time $t$ in
the rest frame of the origin from a location at an axial distance $z$ 
is received by the observer at a time 
$$
T=t-z/c.
\eqno(28)
$$
Our goal is to calculate the specific intensity $I_\nu$ on the surface 
of the shell facing the observer at a distance $R$ from the axis at 
an observer's time $T$, $I_\nu(T,R)$. It depends on the emission and
absorption coefficients at different values of $z$, $j_\nu(T,R,z)$ 
and $\kappa_\nu(T,R,z)$. We need to determine which part of the shell
contributes to the emission and absorption at $(T,R,z)$. In what 
follows, we describe a mathematical procedure for such a determination.

At any given time $t$, the inner and outer radii of the emitting shell 
are known from the blast wave dynamics. It is easy to calculate the time $T$ 
for radiation emitted at a distance $R$ from the axis to arrive at the
observer from equation (28). The possible arrival time fills a band of 
``V'' shape on the $T-t$ diagram for a given distance $R$, as shown in 
Figure~6. Its shape depends on the thickness and dynamics of the shell 
as well as the off-axis distance $R$. One can read off this $T-t$ diagram 
the time $t$ when the emission from the shell at an off-axis distance
$R$ can reach the observer at any given time $T$. Once the time $t$ 
is determined, the axial distance of the emitting region is given by
$$
z=c(t-T).
\eqno(29)
$$
  From equations (18) and (19),  we can determine the comoving energy and 
electron number densities, $U^\prime$ and $n^\prime_e$ (which is equal to
half of the nucleon number density, $n^\prime$, for Wolf-Rayet winds). 
It is then a simple matter
to calculate the emission and absorption coefficients, $j^\prime
_{\nu^\prime}
(T,R,z)$ and $\kappa^\prime_{\nu^\prime}(T,R,z)$, in the comoving frame as a 
function of the comoving frequency $\nu^\prime$. To obtain radiation 
quantities
in the rest frame of the origin, we use the following standard transformation
(e.g., Rybicki \& Lightman 1979):
$$
\nu={\nu^\prime\over \gamma(1-\beta\mu)},
\eqno(30)
$$
$$
j_\nu(T,R,z)={j^\prime_{\nu^\prime}(T,R,z)\over [\gamma(1-\beta\mu)]^2},
\eqno(31)
$$
$$
\kappa_\nu(T,R,z)=\kappa^\prime_{\nu^\prime}[\gamma(1-\beta\mu)],
\eqno(32)
$$
where $\gamma$ and $\beta=(1-\gamma^{-2})^{1/2}$ are the Lorentz factor
and the speed (in units of $c$) of the emitting shell material, and 
$$
\mu={z\over \sqrt{R^2+z^2}}
\eqno(33)
$$
is the cosine of the angle between the symmetry axis and the line passing 
through the origin and the emitting point.

With the emission and absorption coefficients $j_\nu$ and $\kappa_\nu$
determined as a function of axial distance $z$, one can write down the
radiative transfer equation along lines parallel to the axis
$$
{dI_\nu(T,R,z)\over dz}=j_\nu(T,R,z)-\kappa_\nu(T,R,z)I_\nu(T,R,z).
\eqno(34)
$$
With the usual definition of optical depth $d\tau_\nu=-\kappa_\nu dz$ we 
find that the specific intensity at the surface facing the observer 
(where $\tau_\nu=0$) is given by
$$
I_\nu(T,R,\tau_\nu=0)=\int^{\tau_{max,\nu}}_0 S_\nu(T,R,\tau_\nu)
e^{-\tau_\nu}d\tau_\nu
\eqno(35)
$$
where
$$
\tau_{max,\nu}=\int_{-\infty}^{\infty}\kappa_\nu dz
\eqno(36)
$$
is the total optical depth through the shell at an off-axis distance of $R$
while
$$
S_\nu={j_\nu\over \kappa_\nu}
\eqno(37)
$$
is the source function. Finally, we integrate over the surface of the shell
to obtain the total flux at the time $T$ and the frequency $\nu$
$$
F_\nu(T)={2\pi\over d^2}\int_0^{R_{max}}I_\nu(T,R,\tau_\nu=0) R dR.
\eqno(38)
$$ 
With this formalism, we explore numerically various combinations of 
parameters, trying to determine the best fits to the observed radio 
light curves. 

Our procedure was to choose $\epsilon_b$ and $\epsilon_e$ and then to
vary the energy and ambient density until the best fit to 
the data was obtained.
For the same magnetic and electron energy factors $\epsilon_b=\epsilon_e=
0.1$, power index of electron energy distribution $p=2.5$, and source 
distance $d=40$ Mpc as in the last section,  we find that a total energy 
of $E_0=1.2\times 10^{49}$ erg and a value of $n_0 r_0^2=1.2\times 
10^{34}$ cm$^{-1}$ for the external $r^{-2}$-density medium fit the radio 
light curves before the second bumps (around day 24) reasonably well (see 
Figure~7). The value for the external medium corresponds to a stellar 
wind mass 
loss rate of $4.0\times 10^{-7} (V_w/10^3$ km s$^{-1}$) M$_\odot$ yr$^{-1}$. 
Therefore, both the total explosion energy and the wind mass loss rate
are not far from those inferred in the last section. In other words, the 
corrections due to dynamical and relativistic effects are relatively 
modest: they are factors of two or less. 

After about day 24, the observed fluxes increase dramatically at all
four frequencies. These increases are incompatible with the 
predictions of a simple constant-energy blast wave model, as evident
in Figure~7. The predicted fluxes are too small by a factor of
two to three at later times. A resolution of this discrepancy is 
suggested by Figure~4(a), where a sudden increase of energy supply
is implied. We therefore consider the simplest case where the 
total energy of the blast wave increases from its initial value
of $E_0$ to a final value of $E_1$ instantaneously at a time $t_1$
(in the rest frame of the origin). After some experimentation, we
find that the combination of $E_1=3.2\times 10^{49}$ erg and $t_1=
100$ days fits the late time light curves reasonably well (Figure~7). 
The fitting is not perfect, nor is it expected to be, given the 
idealized nature of our model. In calculating the blast wave dynamics
and the radio emission, we have assumed that the promptly injected
energy is shared instantaneously by all material in the uniform
shell, which leads to a jump in the velocity and other properties
of the emitting shell (see Figure~8). The transrelativistic nature
of the emitting region is clear in Figure~8, with a speed 
$\beta=v/c$ between 0.6 to about 0.9  in the time interval of 
interest. It is interesting to note 
that the energy injection occurs about 100 days after the explosion
 in the rest frame of the origin. 
Relativistic effects make its presence felt much earlier in the 
light curves in the observer's frame, around day 22. Also, the jump
in energy is substantial,  to a value that is $\sim 2.6$ times 
higher. 

The inferred total energy and dynamics of the blast wave depend on
the magnetic and electron energy factors. To illustrate the
dependence, we consider the extreme case with a tiny magnetic
energy factor $\epsilon_b=10^{-6}$ and a maximum possible 
electron energy factor $\epsilon_e=1$ \footnote{Strictly speaking,
the choice $\epsilon_e=1$ is not consistent with our blast wave
model which assumes implicitly that most of the energy is 
carried by  nucleons. It should be adequate for a rough parameter
study, however.}. 
These parameters are close to those used by Waxman \& Loeb (1999)
in their subrelativistic model for RSN 1998bw with a thermal electron
distribution.
For a power index of electron energy 
distribution
$p=2.5$ and source distance $d=40$ Mpc as before, we find a
reasonable fit to the observed radio emission (see Figure~9)
for the following set of parameters:  a total initial energy of
$E_0=1.7\times 10^{50}$ erg; a stellar wind mass loss rate of 
$6.2\times 10^{-5} (V_w/10^3$ km s$^{-1}$) M$_\odot$ yr$^{-1}$;
and a jump to a total energy of $E_1=4.5\times 10^{50}$ erg at
a time $t_1=50$ days in the rest frame of the origin. Note that
the rising parts of all four model light curves at days 20 -- 40
are nearly parallel to one another, 
in a better agreement with observations than the case with 
$\epsilon_b=\epsilon_e=0.1$ considered earlier (see Figure~7). The 
reason is 
that the shock speed, also shown in Figure~8, is significantly 
lower in the $\epsilon_b=10^{-6}$ case than in the $\epsilon_b
=0.1$ case in the time interval of interest; a lower shock speed 
tends to steepen the rise in the radio fluxes at short, optically 
thin wavelengths, since a larger portion of the re-energized
shell contributes to the total emission at the given time.  
Compared to the model of Waxman \& Loeb (1999), which was primarily
intended for the observations on day 12,  our model has a higher shock
velocity ($0.6c$ vs. $0.3c$) and a higher energy by a factor of a few. 
The differences may be related to the fact that our model has a
decelerating shock wave, includes relativistic effects, and
assumes a power law electron distribution.

These models show that models with a wide range of $\epsilon_b$
can approximately represent the data.
The scaling between these models is discussed in the next section.
However, we have found that acceptable models are even more widely
distributed and can be found with $\epsilon_e=0.01$.
It appears that changes in $\nu_m$ and changes in the time lag
effects can cancel each other out.
The energy and ambient density of these models remains in the range
of those given above.

We have also investigated models in which the flux rise is due to
a jump in the ambient density instead of an energy increase.
Although it is possible to obtain an increase in the extrapolated flux
going to late times, it is not possible to obtain a fit during the
transition period (20-40 days).
In particular, the fluxes remain flat during this period instead of rising.

In the models with low $\epsilon_b$ so that radiative losses are
less important, the predicted optical emission 
is comparable to that observed in day 1 of SN 1998bw (Galama et al. 
1998b). It points to the intriguing possibility that 
 the rapidly-fading optical transient of GRB 980425 was
detected on day 1. 
This possibility is diminished, however, by the fact that the model 
does not fit the early (less complete) radio light curves, before 
about day 10.

\section{POWER LAW DYNAMICS}

Many of the previous discussions of GRB afterglow 
light curves have assumed power
law evolution based on smooth properties of the ejecta and the
surrounding medium (M\'esz\'aros, Rees,  \& Wijers 1998 and references therein).
RSN 1998bw shows a more complex evolution, but our model indicates
that over substantial periods of time the evolution is described by
a constant energy shell in a $\rho\propto r^{-2}$ medium.
During these times the radio flux evolution is described by
$F_{\nu}\propto t^{-1.6}$ at optically thin wavelengths.
M\'esz\'aros, Rees,  \& Wijers (1998) discuss the expected power law
evolution in a $\rho\propto r^{-2}$ medium for a relativistic flow
and find that for $\nu >\nu_m$ and constant energy, $F_{\nu}\propto
t^{(1-3p)/4}$.
For $p=2.5$, we have $F_{\nu}\propto t^{-1.62}$, in good agreement with
the observations of RSN 1998bw.
The observed time period 12--24 days is especially useful 
because self-absorption and $\nu_m$ affect the evolution and provide
diagnostic information.
The optical depth at 1.4 GHz is moderately high and we expect
$F_{\nu}\propto r^2\propto t$, which is in agreement with the observations
(Fig. 1).

The fact that a constant energy explosion in a $\rho\propto r^{-2}$ medium
describes the $t=12-24$ day evolution can be used to find scaling relations
for the acceptable models.
The model parameters are $E$, $A$ (where $\rho =Ar^{-2}$), $\epsilon_e$,
and $\epsilon_b$.
The parameter $p$ is simply determined by the observations.
The observational constraints on these parameters can be described by
$\nu_m$, $F_{\nu_m}$, and $\nu_A$, where $\nu_A$ is the frequency at which the
spectrum becomes self-absorbed.
The expected evolution of these quantities for a relativistic explosion
in an $r^{-2}$ medium is (cf. Waxman 1997; M\'esz\'aros et al. 1998)
$$
\nu_m\propto\epsilon_e^2\epsilon_b^{1/2}E^{1/2}t^{-3/2},
\eqno(39)
$$
$$
F_{\nu_m}\propto\epsilon_b^{1/2}AE^{1/2}t^{-1/2},
\eqno(40)
$$
$$
\nu_A\propto\epsilon_e^{-1}\epsilon_b^{1/5}A^{6/5}E^{-2/5}t^{-3/5}.
\eqno(41)
$$
There are 3 constraints on the 4 model parameters, so there is not expected
to be a unique model fit to the data.
One way of describing the scaling of acceptable models is to fix
$\nu_m$, $F_{\nu_m}$, and $\nu_A$ at some time $t$ and find how 3 of
the model parameters depend on the fourth one.
For example, we find
$$
E\propto\epsilon_b^{-1/5},\qquad A\propto\epsilon_b^{-2/5},\qquad
\epsilon_e\propto\epsilon_b^{-1/5}.
\eqno(42)
$$
If one dynamical model with $\epsilon_b=0.1$ and $\epsilon_e=0.1$ fits
the data, then another model with $\epsilon_b=10^{-6}$, $\epsilon_e\sim 1$,
$E$ up by 10, and $A$ up by 100 should fit the data equally well.
This is in approximate accord with the results in the previous section.
In the opposite direction, increasing $\epsilon_b$ to equipartition
with the postshock gas ($\epsilon_b=0.5$) leads to moderately small decreases
in the other parameters.
Our detailed models, discussed in the previous section, show that the
range of acceptable parameters is broader.
The emission is not from a single region, as is assumed in the power law
model, but is integrated over a shell with time lags.
These conditions allow a greater variety of models.

For the flux evolution in the optically thin regime, equations (39) and (40)
can be combined to yield
$$
F_{\nu}\propto \epsilon_e^{p-1}\epsilon_b^{(p+1)/4}AE^{(p+1)/4}t^{(1-3p)/4}.
\eqno(43)
$$
The observed flux increase from before day 24 to after day 32 is a factor
of 2.2 (see \S~2).
If this increase is due to an energy increase, equation (43) shows that 
$E\propto F_{\nu}^{4/(p+1)}$ and so $E$ must increase by a factor of 2.5 for
$p=2.5$.
 This result agrees well with the factor of 2.6 found in
the detailed models.
The failure of a density increase to account for the flux rise can
also be seen from the power law dynamics.
The flux at an optically thick wavelength is expected to increase
as $r^2$.
A boost in the shock energy accelerates the expansion rate and accelerates
the flux increase.
An increase in the ambient density decelerates the shock and the flux increase.
The data at 1.4 GHz, an optically thick
wavelength, show an accelerating rise through the time of the flux increase.
The implication is that the increase is due to an energy, not a surrounding
density, increase, which is in accord with our detailed models.

An extrapolation of the constant $E$ evolution to day 10 and earlier
yields fluxes at  8.6 GHz that are higher than those observed.
Equation (41) shows that an attempt to model this solely as an energy
jump yields a higher value of $\nu_A$ at early times.
This would imply that the observed spectrum should be in the self-absorbed
regime, or $F_{\nu}\propto \nu^2$.
The observed spectrum is not compatible with this form, which shows that the
early flux increase requires factors in addition to or other than
an energy increase.

\section{RSN 1998bw AND GRB 980425}

Our  models have implications for the nature of RSN 1998bw in
the context of SN 1998bw and the possibly associated GRB 980425.
Models for the light curve and some spectra of SN 1998bw indicate that
the progenitor star was the heavy element core of a massive star with
a final mass of $13.8\Msun$ (Iwamoto et al. 1998) or $6\Msun$
(Woosley et al. 1999).
Such stars are observed as Wolf-Rayet stars, which have winds with velocities
in the $V_w=1,000-2,500\kms$ range and mass loss rates $\dot M=
10^{-5}-10^{-4}\ml$  (Willis 1991).
Our  $\epsilon_b=0.1$ model for RSN 1998bw implies $\dot M/V_w
\approx (4\times 10^{-7}
\ml)/(1,000\kms)$  and the  $\epsilon_b=10^{-6}$ model implies
$\dot M/V_w\approx (6\times 10^{-5}\ml)/(1,000\kms)$.
The density range in the models  overlaps
 the circumstellar density expected for a Wolf-Rayet
star; the high $\epsilon_b$ models yield a density that is somewhat 
too low,
so there is a preference for the low $\epsilon_b$ models. The 
low $\epsilon_b$ models also appear to fit the rise in the radio 
light curves at days 20 -- 40 better. 

The blast wave energies estimated during the observed period 
beginning on day 12
 range from $1.2\times 10^{49}$ ergs with $\beta=v/c\gsim 0.8$ for
the $\epsilon_b=0.1$ model and $1.7\times 10^{50}$ ergs with 
$\beta\gsim 0.5$ for
the $\epsilon_b=10^{-6}$ model.
Immediately after the impulsive energy increase, 
the energies are $3.2\times 10^{49}$ ergs and $4.5\times 10^{50}$ ergs
with $\beta\approx 0.85$ and $\beta\approx 0.6$ 
 for the high and low $\epsilon_b$
models, respectively.
These energies can be compared to those expected in the shock-accelerated
high velocity ejecta of SN 1998bw.
There is the possibility that the high radio luminosity is simply
because of the higher explosion energy than that normally found in
Type Ib/c supernovae.
Woosley et al. (1999) estimate that their spherical model 
for SN 1998bw produces $5\times 10^{49}$
ergs of ejecta with $\beta=c/3$.
Although their models are nonrelativistic,
Matzner \& McKee (1999) estimate the relativistic mass ejection that can be
expected from an explosion like SN 1998bw.
They note that a stripped massive star may not be sufficiently compact to
produce any relativistic ejecta.
A $3\times 10^{52}$ erg explosion with $6\Msun$ of ejecta in a sufficiently
compact star should yield $\sim (1-2)\times 10^{48}$ ergs of relativistic
ejecta.
The shock acceleration process yields a mass (and energy) that declines
steeply with velocity, so the results are sensitive to the details of
the model.
The energy inferred for RSN 1998bw appears to be  too high
for the shock accelerated ejecta of SN 1998bw in spherical models, 
but this result is
not conclusive.

Our model also places constraints on the time dependence of the energy
input to the blast wave.
For $t>12$ days, we find that the blast wave evolves with approximate
constant energy except for an episode during which the energy is
increased by a factor $\sim 2.5$.
The change does not appear to be due to the circumstellar medium and
is probably due to  inner material that catches up
with the interaction region.
This type of evolution is not expected for energy injection from 
shock-accelerated, high velocity supernova ejecta, which presumably
have a smooth distribution.
This property, along with the high energy in RSN 1998bw, suggests
the presence of a central engine in the explosion.
The fact that RSN 1998bw was accompanied by a Type Ic supernova with
$\sim 10\Msun$ of material moving at moderate velocity (Galama et al. 1998b;
Iwamoto et al. 1998), implies that the high velocity outflow giving rise
to RSN 1998bw is not spherically symmetric, but occupies only a fraction
of the total $4\pi$ solid angle if it originated in a central engine.
MacFadyen \& Woosley (1999) have proposed a model for SN 1998bw in which
collimated, high-energy flows are created by a disk around a central
black hole.
The success of spherically symmetric models in describing normal GRBs
is usually attributed to relativistic beaming; only a small part
of a possibly asymmetric source is being observed.
This argument does not apply to our model for RSN 1998bw because
the velocities are only mildly relativistic and it is thus
surprising that a spherically symmetric model is so successful.
The introduction of asymmetries in the models brings in new parameters
and we have not investigated such models.

The type of evolution that is observed in RSN 1998bw may be related to that
observed in the optical afterglow of GRB 970508.
In that case, there is evidence for a steady or decreasing flux
before the 1.5 day rise phase leading to the maximum (Pedersen et al. 1998).
An extrapolation of the early evolution implies that
the rise is by a factor of 4.
M\'esza\'ros, Rees, \& Wijers (1998) suggest that the rise is due
to a multicomponent flow or to a lower $\Gamma$ shell catching up with
the main shock front.
Both of these models have been calculated in more detail by
Panaitescu, M\'esza\'ros, and Rees (1998).
In their model with late impulsive energy input, the energy
increases from $0.6\times 10^{52}$ ergs to $2.4\times 10^{52}$ ergs.
Although the energies are larger than those inferred in RSN 1998bw,
the factor increase in energy is  comparable to
 the one that we have advocated for RSN 1998bw.
 A multicomponent flow model does not appear plausible for RSN 1998bw
 because of the continuity of the radio spectral indices during the time
 of the flux increase.

The fact that we have found that RSN 1998bw has a central engine related
to those found in normal GRBs strengthens the association with
GRB 980425 and suggests that the relation of the radio emission to
the early $\gamma$-ray emission be investigated.
However, there are several reasons why an energy conserving synchrotron model
cannot be simply extrapolated back to the very early times (cf. Iwamoto 1999).
First, we have found evidence in the radio light curves for at least one 
episode
of impulsive energy injection.
Additional episodes could have also occurred before day 3 when the radio
observations were initiated.
The result would be lower flux values  at early times.
Second, synchrotron losses become more important at early times.
The frequency at which the synchrotron loss time equals the age is
$\nu_c\propto t^{1/2}$ for constant energy evolution in an $r^{-2}$ medium.
The peak frequency evolves as $\nu_m\propto t^{-3/2}$, so that the
blast wave was radiative in the past.
Taking $\nu_m=5.6$ GHz on day 10  implies radiative evolution
before $t\sim 0.1$ day in our high $\epsilon_b$ model.
Cooling is less important with low $\epsilon_b$.
Third, the unusual behavior observed between days 3 and 12 may indicate
that it may not be possible to extrapolate our model to small radii.
Finally, the early evolution may be affected by the
character of the initial injection of
energy.
If the shell has an initial Lorentz factor $\Gamma_o$, its mass is
$M=E_o/\Gamma_o c^2$, where $E_o$ is the initial energy.
The deceleration radius, $r_{\rm dec}$, occurs when the shell has swept
up a mass $\Gamma_o M$ in the circumburst environment.
Thus
%\begin{equation}
$$
r_{\rm dec}={E_o\over \Gamma_o^2 c^2 (\dot M/V_w)}
=2\times 10^{12} E_{o48}\Gamma_{o1}^{-2}
\left(\dot M/V_w\over 6.3\times 10^{12}
{\rm~ g~cm^{-1}}\right)^{-1}\quad {\rm cm},
\eqno(44)
$$
%\end{equation}
where $E_{o48}=E_o/(10^{48}{\rm~ergs})$, $\Gamma_{o1}=\Gamma_{o}/10$,
and the reference value of $\dot M/V_w$ corresponds to
$\dot M/V_w = (1\times 10^{-5}\ml)/(1,000\kms)$.
The energy $E_o=10^{48}$ ergs is a lower limit based on the radiated
energy in $\gamma$-rays.
The shell radius at $t=10$ s is $r\approx 2\Gamma_o^2 ct=6\times 10^{13}
\Gamma_{o1}^2$ cm.
If $\Gamma_o$ starts at the lower end of the plausible range or if
$E_o$ is high, the burst observations may be during the initial deceleration
period.
Other possible factors are electron scattering opacity and
$\gamma-\gamma$ opacity for the GRB.
If the electron scattering opacity is 0.4 cm$^2$ g$^{-1}$, as expected
for a baryon dominated plasma, electron scattering is not important.
For the relatively low energy burst considered here, $\gamma-\gamma$ opacity
is not important (cf. Piran 1996).

\section{CONCLUSIONS}

We have modeled the radio emission from SN 1998bw as the synchrotron
emission from a rapidly expanding shock wave.
 We assumed that the electron energy and the magnetic energy densities are
constant fractions of the postshock energy density, as is commonly done
in models of both radio supernovae and GRB afterglows.
The result of our modeling is that the radio evolution is consistent
with expansion into a $\rho\propto r^{-2}$ circumstellar wind.
The estimated wind density is consistent with that expected around
the progenitor star that gave rise to the optical SN 1998bw and with
the type of environment commonly deduced around radio supernovae.
However, the radio light curves show a period of flux increase
that appears to require an episode of impulsive energy injection.
The evolution at other times is consistent 
with constant blast wave energy.
We speculate that the freely expanding explosion ejecta are nonuniform and that
a shell caught up with and energized the blast wave.

Katz (1999) has recently proposed a different model for RSN 1998bw
in which the power for the source is provided by radioactivity.
The $\gamma$-rays lines from radioactivity Compton scatter electrons
which then move out in a mildly relativistic flow.
Possible problems for the model are that the $^{56}$Ni produced
in the explosion must be mixed to the surface of the debris and the
origin of the energetic electrons needed for the synchrotron emission
is unclear.

The association of SN 1998bw with GRB 980425 appears likely based not
only on the spatial and temporal coincidence of the events (Galama
et al. 1998b), but also on the unusually high energies of the optical
and radio supernova phenomena.
The  finding that the ejecta are nonuniform gives
additional evidence for a relation between the two events because
the ejecta in normal radio supernovae are inferred to be approximately
smoothly distributed, as expected for shock-accelerated ejecta.
The optical light curve of the GRB 970508
afterglow may indicate discrete structure in the ejecta; in other GRB cases,
the evolution can be described by a constant energy model as is also
inferred for substantial parts of the evolution of RSN 1998bw.
Although the environment of RSN 1998bw is similar to that inferred
around normal radio supernovae, the explosion event shows a link
to the more energetic GRBs.
Woosley et al. (1999) and MacFadyen \& Woosley (1999) have also 
linked SN 1998bw to the GRBs based on the high energy of the event.
The outflow may be powered
 by neutrino-antineutrino annihilations soon after the
formation of a central black hole
or by rotational energy extracted electromagnetically 
from a Kerr black hole or its
nearby accretion disk
 (MacFadyen \& Woosley 1999; Paczy\'nski 1998).

Cosmological GRBs with well-observed afterglows, including
GRB 970228 (Wijers et al. 1997) and GRB 970508 (Waxman 1997;
Granot, Piran, \& Sari 1998),
have been inferred to be expanding into a medium with a low, constant
density ($n_o\approx 1$ cm$^{-3}$).
Power law declines are observed, but with a time dependence
$F_{\nu}\propto t^{-w}$ with $w=1.10-1.20$.
The steeper decline observed in RSN 1998bw ($w=1.6$) is an important
part of the evidence for expansion in a $\rho\propto r^{-2}$ medium.
The steeper decline is also expected for the optical and X-ray afterglows,
which can account for the unusual lack of an X-ray afterglow if
SN 1998bw is associated with GRB 980425.
The expected early radiative losses are another factor (Bloom et al. 1998).

Bloom et al. (1998) have discussed the notion that the SN 1998bw/GRB 980425
event is a prototype of a new class of GRB associated with supernovae
(S-GRB).
One of the attractive features of the association was that the single pulse
nature of GRB 980425 could be attributed to the shock wave generated by
the smooth, shock-accelerated ejecta, which are not expected to contain
internal shocks.
In our model, the radio emission is driven by ejecta with considerable
structure so that the association with a single pulse GRB may not
be significant.
However, our work has brought out another suggestive distinction between the
S-GRB and normal GRBs.
RSN 1998bw shows evidence for interaction with a circumstellar wind,
as expected for its massive star progenitor.
In the cases where there are sufficient data to make a determination,
the observations of normal GRBs are consistent with interaction with
a medium with a low, constant density.
The situation is not expected in the immediate vicinity of a massive
star and may point to a different kind of progenitor object in these cases.

\acknowledgments
We are grateful to Dale Frail and Shri Kulkarni for information on
their radio observations and their comments, and to Claes Fransson,
Eli Waxman, Stan Woosley, and the referee, Jonathan Katz, for
useful comments on the manuscript.
Support for this work was provided in part by NASA grant NAG5-8232.

\clearpage

\begin{figure}

\caption{The full observed radio fluxes and the nonrelativistic model
fits for day 12 and later. As explained in \S~3, the flux at 6 cm is 
fitted analytically. Together with the model spectral indices shown 
in Figure~2, it yields the model fluxes at other wavelengths.}

\caption{Spectral indices of radio emission at day 12 and later. 
The solid lines are model fits using a simple synchrotron 
self-absorption model with electrons in a power-law energy 
distribution truncated from below. The fitted quantities are
shown in equation (7).}

\caption{The peak spectral luminosity vs the product of the time
of the peak and the frequency of the measurement, 
based on Fig. 4 of Chevalier (1998) with
the addition of SN 1998bw.
The observed supernovae are designated by the last two digits of the
year and the letter, and are of Types II ({\it stars}), Ib ({\it squares}),
and Ic ({\it crosses}).
%SN 1983N has two points to indicate a possible range and SN 1981K and
%SN 1984L have lines to indicate that the plotted points are limits.
The dashed lines show the mean velocity of the radio shell if
synchrotron self-absorption is responsible for the flux peak;
a value $p=2.5$ is assumed.  }

\caption{Deduced quantities of a uniform, nonrelativistically moving
shell between day 12 and day 80 from model fits to the observed 
spectral indices and the 6 cm flux. Shown are (a) the time evolution
of the total energy, (b) the evolution of
the shock radius, (c) the electron density
in the emitting shell as a function of radius, and (d) the ratio of 
the mass loss rate and speed of the progenitor's wind.}

\caption{The coordinate system for light travel time effects. Note 
that the emitting shell is marked by a ring and that $R$ denotes 
the distance of the line of sight away from the axis and $z$ the 
distance along the line of sight.}

\caption{The conversion diagram for the arrival time $T$ at the 
observer and the time $t$ in the rest frame of the origin of
emission from a relativistically moving shell at a given 
off-axis distance $R$. The four points, A-D, correspond to
the four points in Figure~5 schematically. }

\caption{Model fits to the observed radio fluxes including dynamical 
and relativistic effects. Constant energy factors $\epsilon_b
=\epsilon_e=0.1$ are assumed. The second bumps in the model light 
curves
are due to a sudden jump in the total energy (to about $2.6$ times
its initial value).}

\caption{Time evolution of the shock speed. The jump is caused by
the jump in the total energy. The transrelativistic nature of 
the blast wave is evident.}

\caption{Model fits to the observed radio fluxes including dynamical 
and relativistic effects in the case of a magnetic energy factor 
$\epsilon_b=10^{-6}$ and an electron energy factor $\epsilon_e=1$. 
See Figure~7 for comparison.} 

\end{figure}

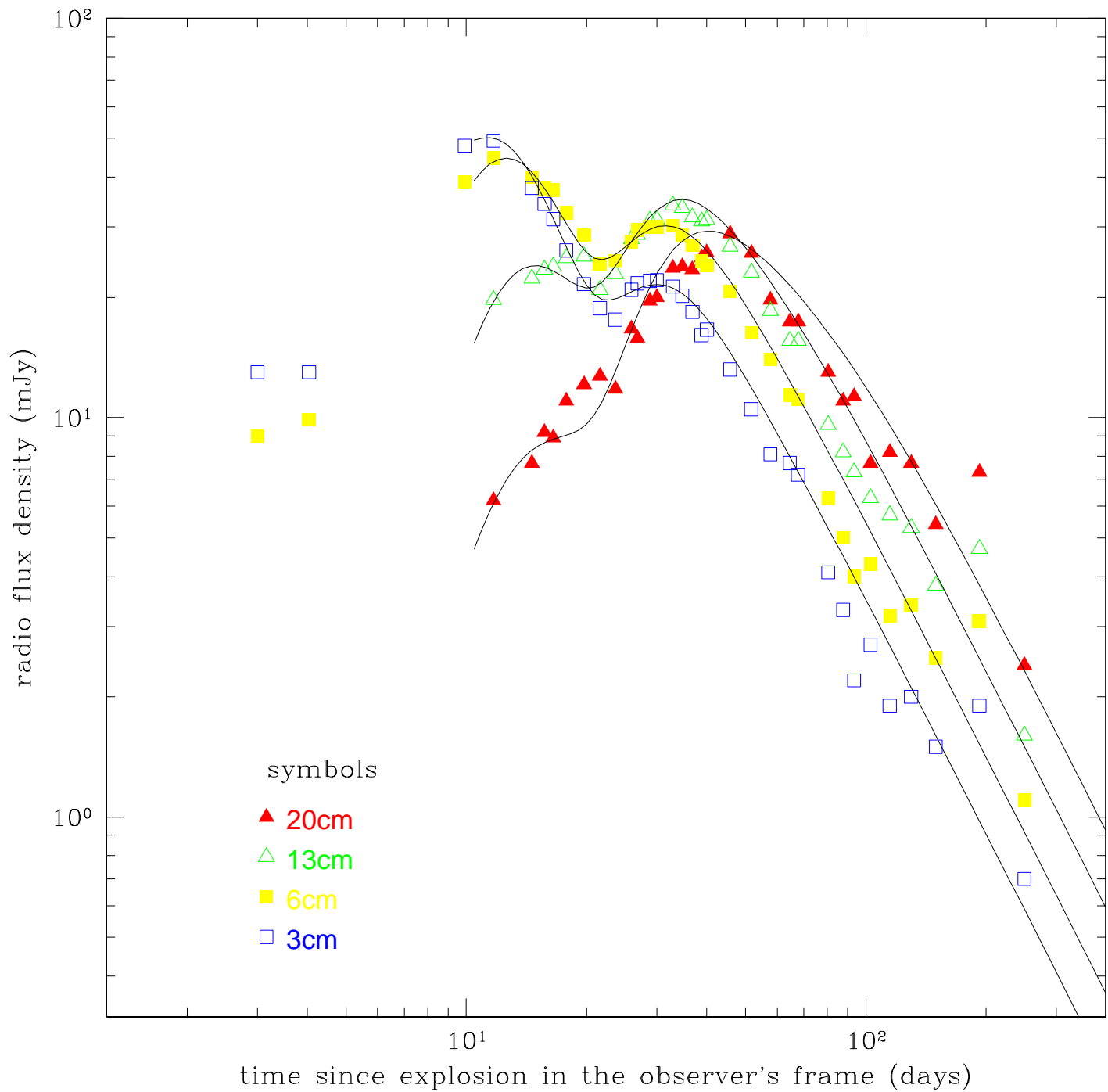
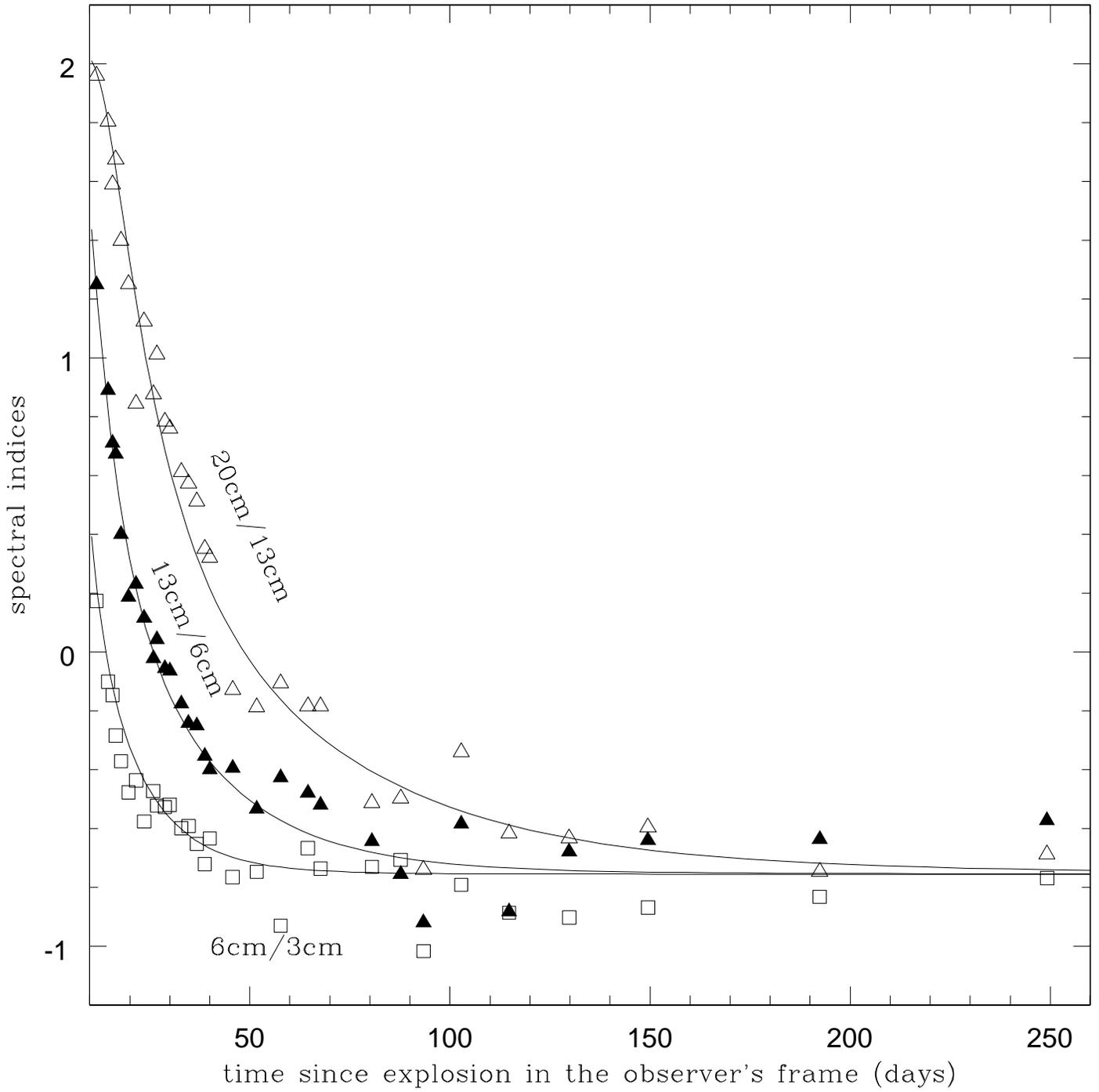
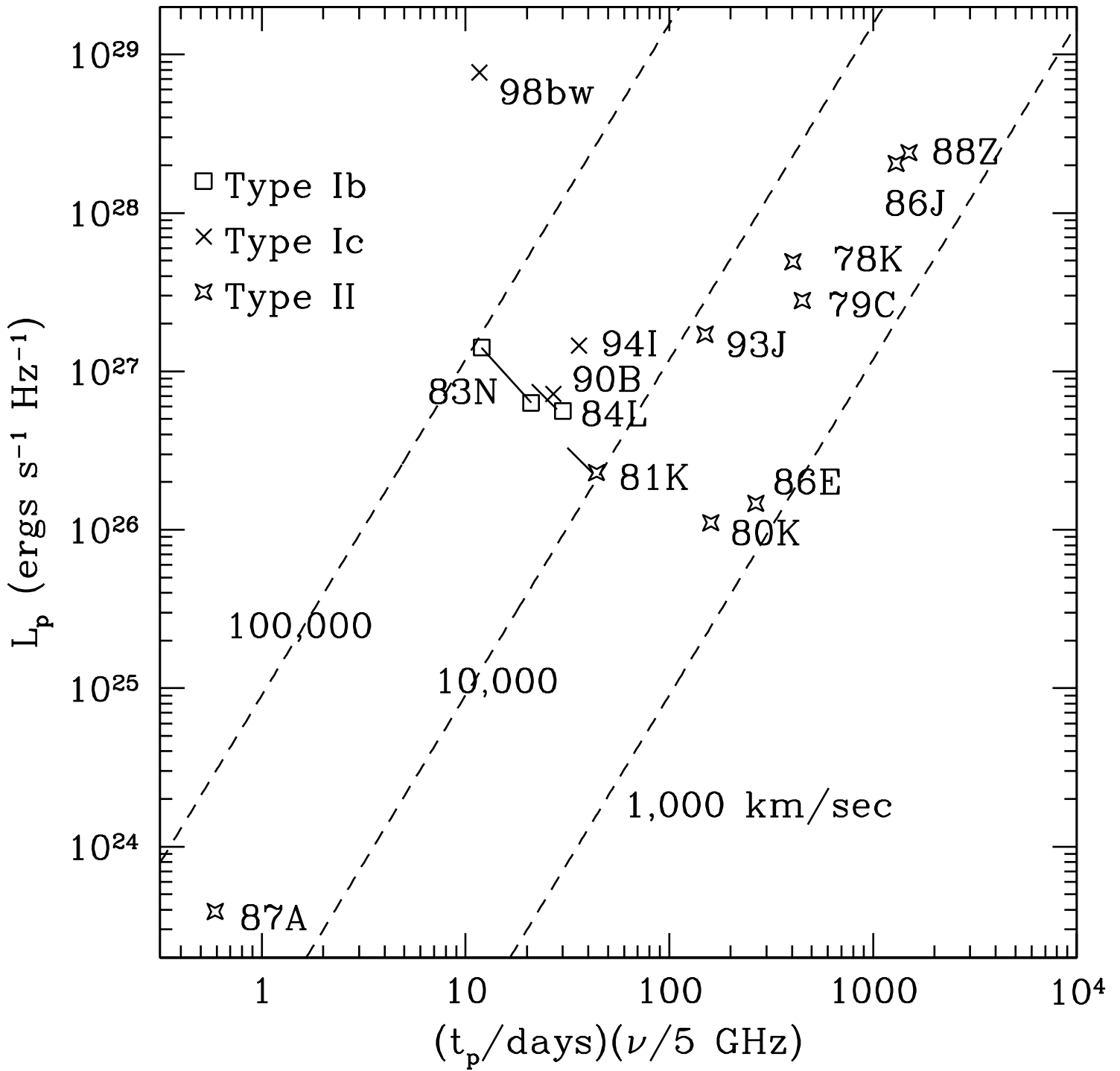
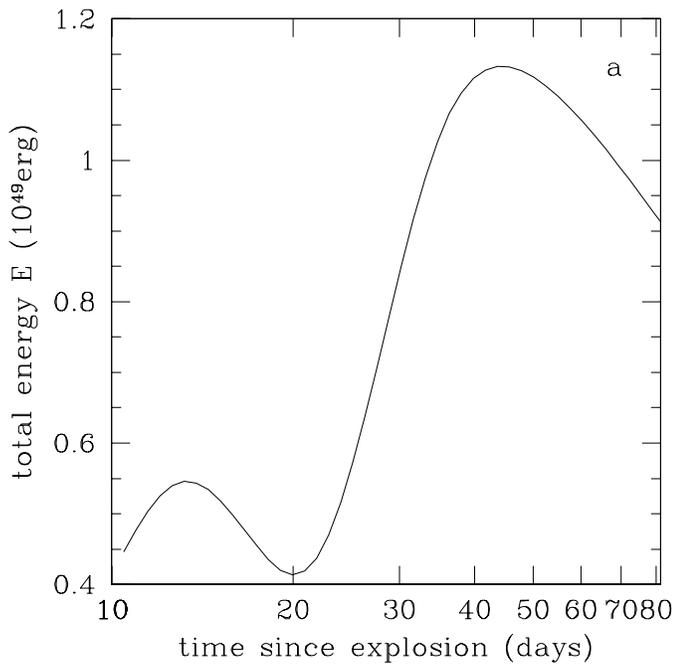
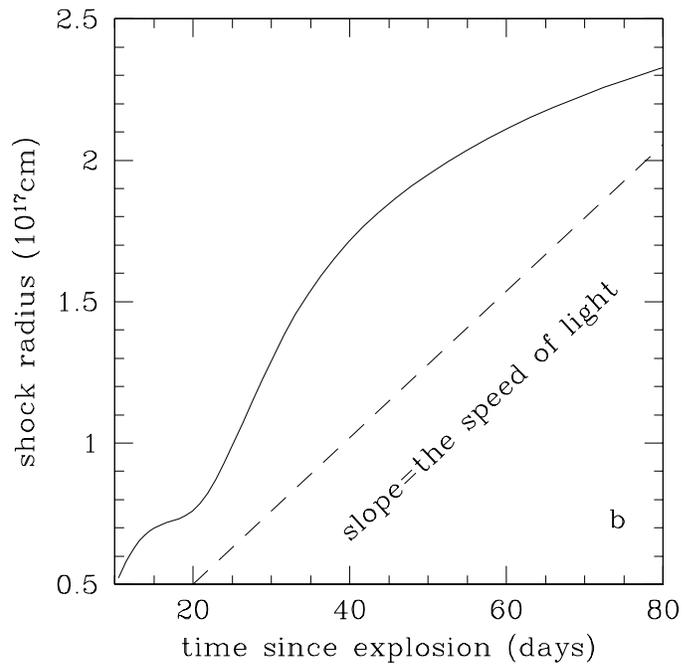
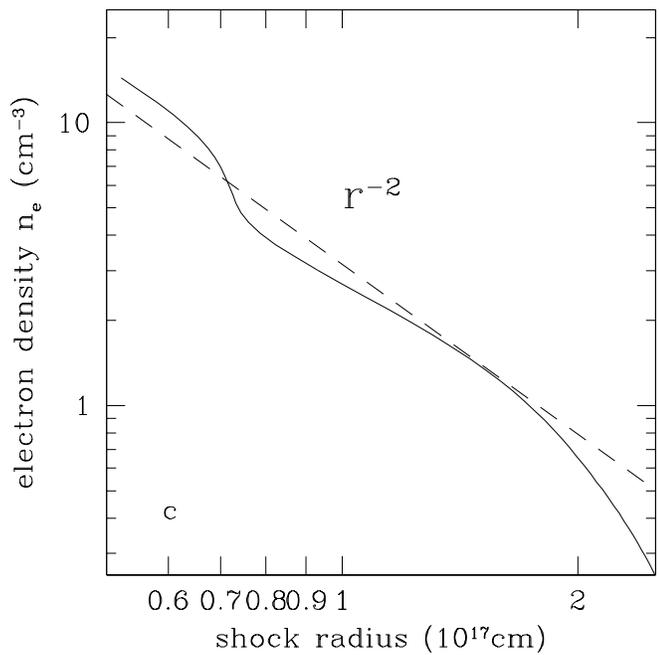
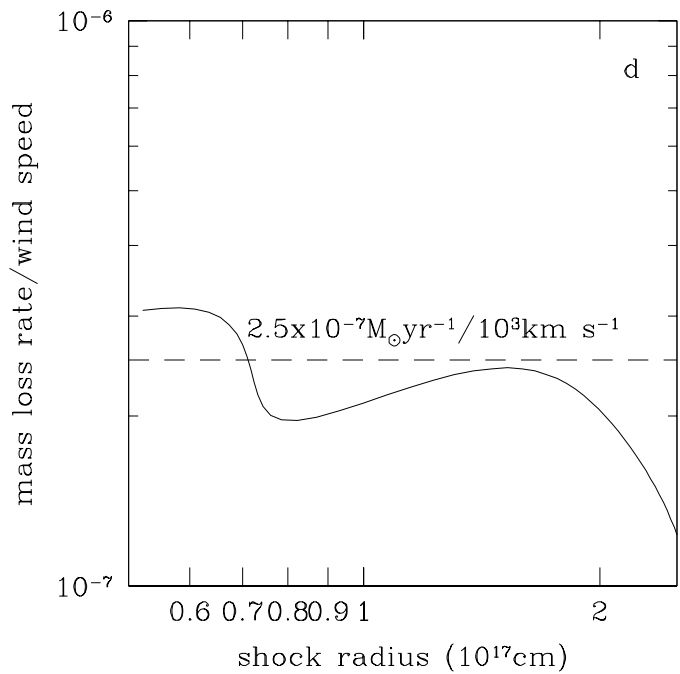
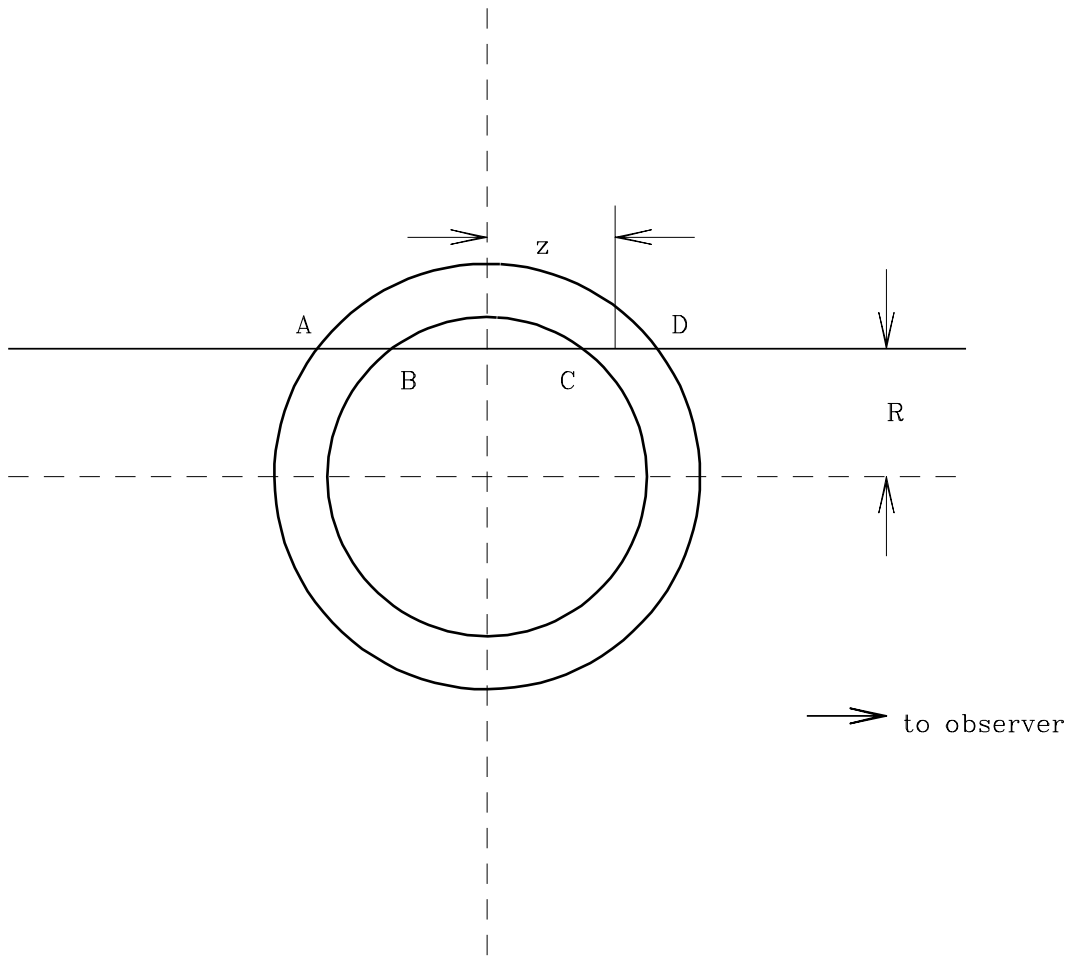
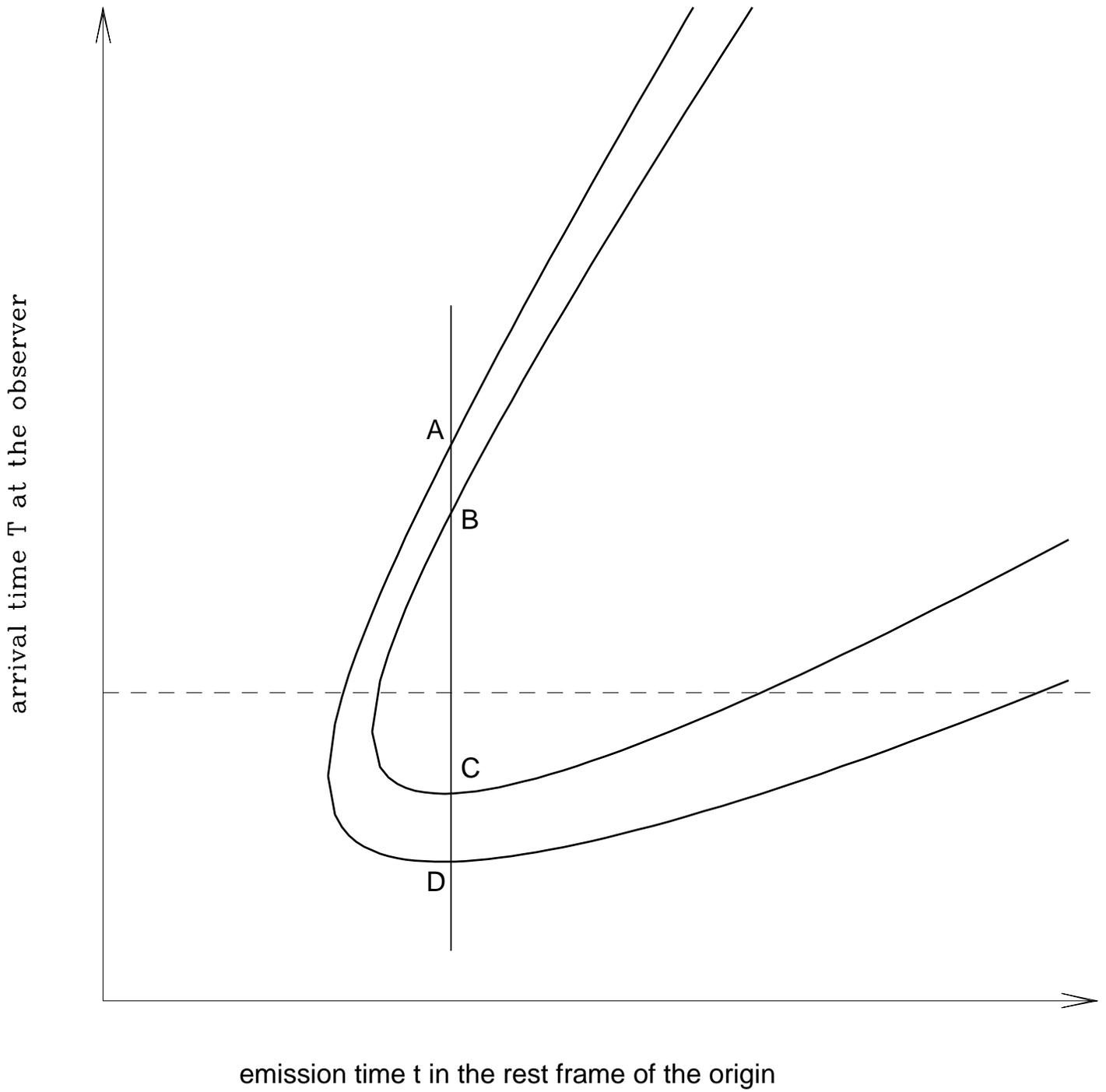
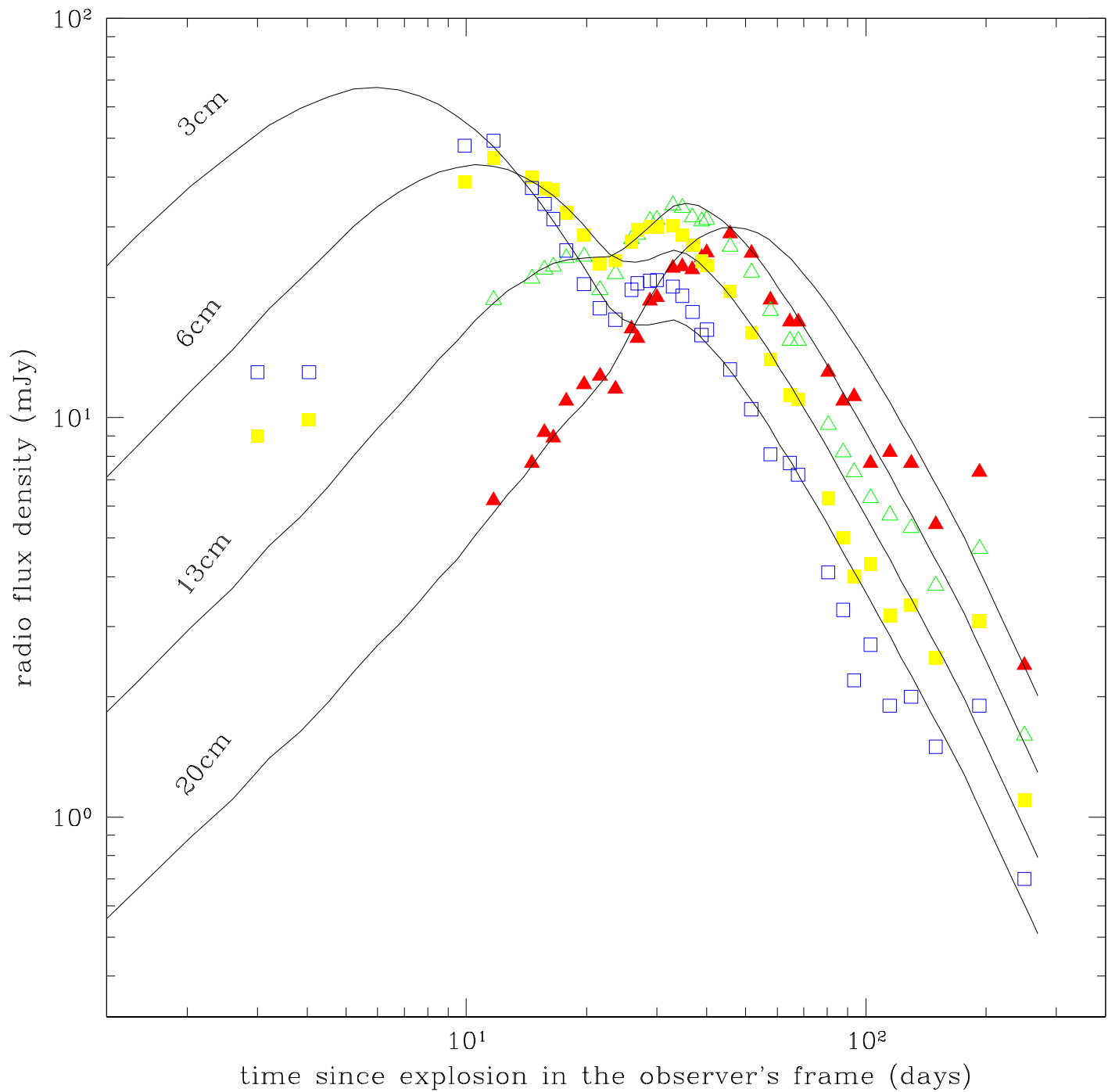
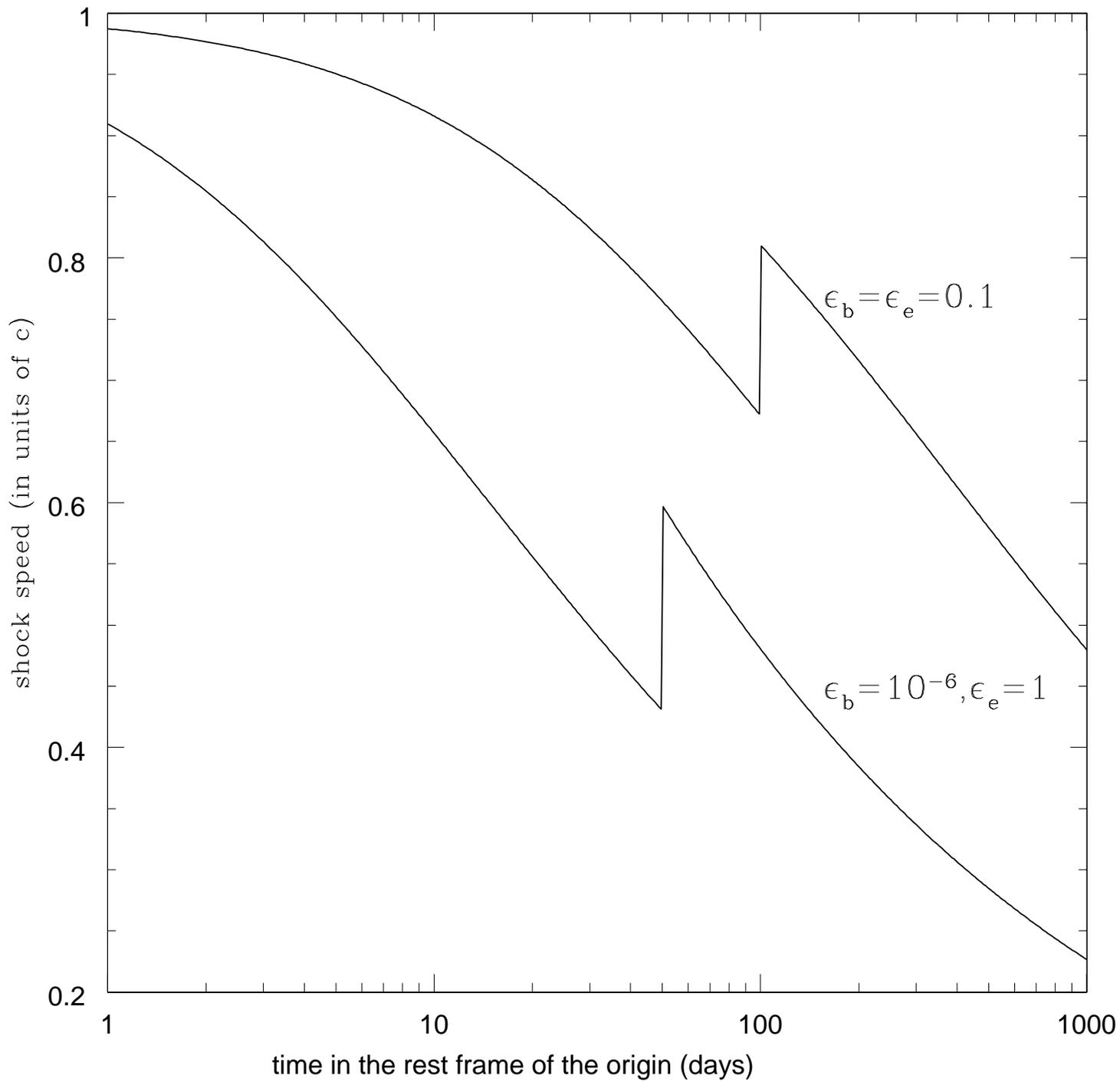
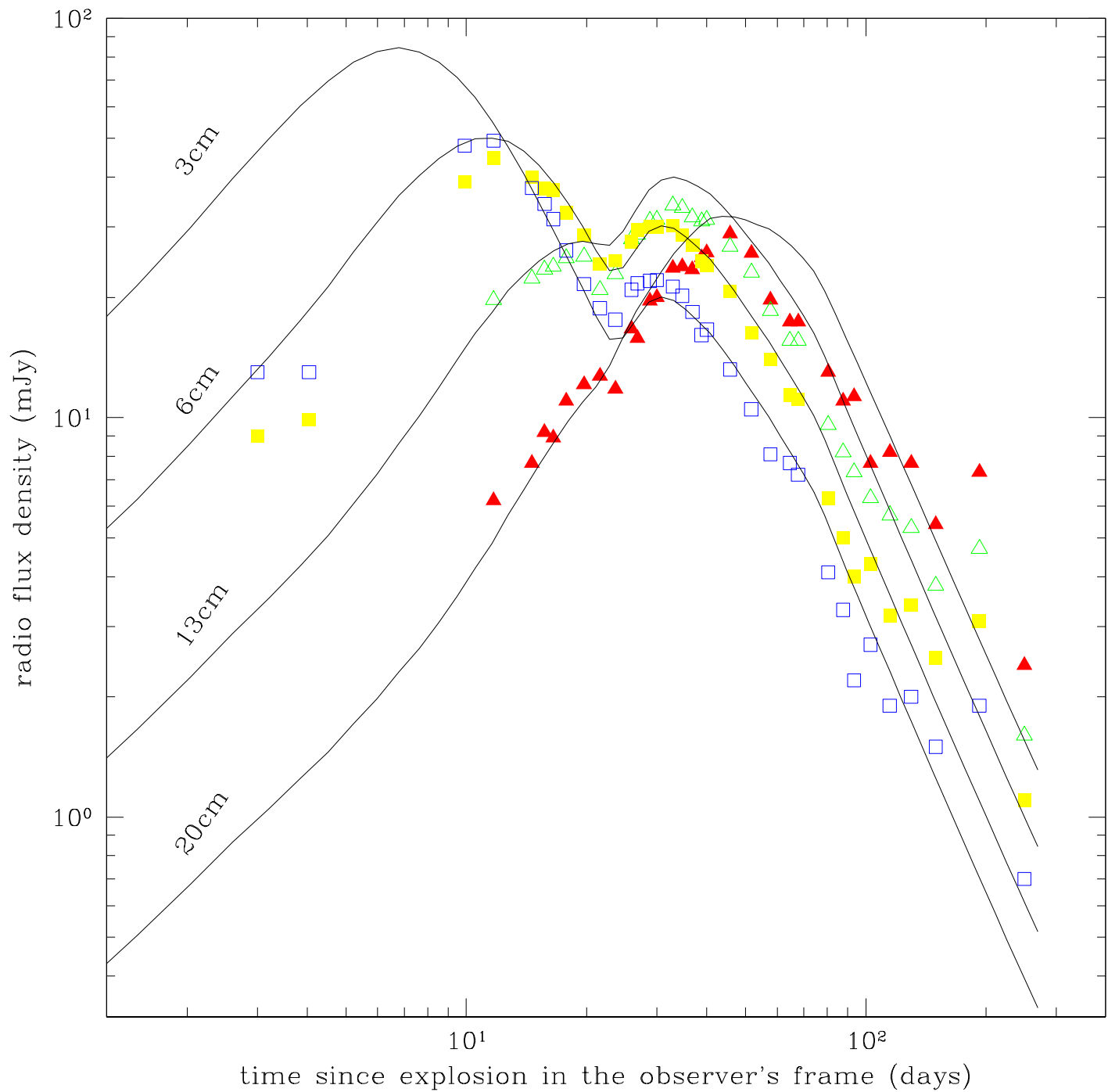
\end{document}